\documentclass[twocolumn,showpacs,showkeys,a4paper,9pt]{revtex4}
\usepackage{amsthm}
\usepackage{mathrsfs}
\usepackage{graphicx}
\usepackage{dcolumn}
\usepackage{bm}
\usepackage{subfigure}
\begin{document}

\markboth{Wen-An Li}
{Simplified proposal for realizing multiqubit tunable phase gate in circuit QED}

\linespread{1.6}

\title{Simplified proposal for realizing multiqubit tunable phase gate in circuit QED}

\author{Wen-An Li\footnote{E-mail: liwenan@126.com} and Yuan Chen}

\address{Department of Physics, School of Physics and Electronic Engineering, Guangzhou University, Guangzhou 510006, China}

\begin{abstract}
We propose a scheme to realize multiqubit tunable phase gate in a circuit QED setup where two resonators each coupling with a qudit are interconnected to a common qudit ($d=4$). In this proposal, only two levels of each qudit serve as the logical states and other two levels are used for the gate realization. The proposal is efficient and simple because only a classical microwave pulse is needed, no matter how many qudits are involved, which significantly reduces experimental difficulty. In non-resonant case, the tunable phase gate can be achieved readily, while under the resonant condition a $\pi$-phase gate can be realized after a full cycle of Rabi oscillation where the gate speed is rather fast due to the resonant interaction. We have shown that the resulting effective dynamics allows for the creation of high fidelity phase gate. The influence of various decoherence processes such as the decay of the resonator mode, and the relaxation of the qudits is investigated. Moreover, the proposed scheme can be easily generalized to realize $N$-qubit phase gate.
\end{abstract}


\maketitle
\section{introduction}
Quantum computer holds promise that it owns the great power to solve classically intractable problems such as factoring a number~\cite{Shor} and searching a data in an array~\cite{Grover}. 
This is, in general, accomplished by performing specific unitary transformations on a set of quantum bits followed by measurement. The basic element of a computer is the logic gate, either in a classical computer or a quantum computer. Now it has been shown that one-qubit gates and two qubit controlled phase gates are universal for constructing a quantum computer, i.e. any multiqubit gates can be achieved by choosing appropriate set of these elementary gates. In practical quantum computing, the implementation of quantum algorithms and quantum error-correction protocols may involve multiqubit quantum gates~\cite{3,4}. As the number of the qubits increases, the procedure of decomposing multiqubit gates into several basic elementary gates becomes more and more complicated. It is necessary to develop a way to realize the multiqubit quantum gate directly. 

In particular, the multiqubit controlled phase gate which shifts the phase of only one of the state components is of great importance. This gate can be widely used in quantum algorithms~\cite{5}, quantum Fourier transform~\cite{6}, and quantum error correction ~\cite{3,7,8,9}. A number of theoretical schemes~\cite{10,11,12,13,14,15} have been proposed to implement the there-qubit quantum gate, and it has been demonstrated experimentally in nuclear magnetic resonance~\cite{9}, linear optics~\cite{16}, ion traps~\cite{17}, and circuit QED systems~\cite{7,18}. However, the controlled phase gates involving more than three qubits have not been experimentally implemented. Though an n-qubit controlled phase gate could be decomposed into the elementary one- and two-qubit gates, it requires much longer times and yields lower overall fidelities. For example, the Toffoli gate implemented with only single- and two-qubit gates requires six controlled-NOT gates and ten single-qubit operations~\cite{19}. Therefore, it is hard to realize the phase gates involving more than three qubits in any system owing to current limits on coherence.

Recently, several schemes, such as $n$-control qubits acting on one target qubit~\cite{15,20,21,22,23,24}, one control qubit simultaneously controlling n target qubits, based on cavity QED or circuit QED~\cite{25,26,27,28}, have been proposed.  For example, Yang et al~\cite{20} present a way to realize an $n$-qubit controlled phase gate with superconducting quantum-interference devices (SQUIDs) by coupling them to a superconducting resonator. The  implementation of three-qubit phase gate requires seven operational steps, and adjusting the level spacings of the SQUID to couple corresponding energy level. In their proposal, the required steps for the $n$-qubit controlled phase gate is $2n+1$, which make the experimental procedure much complicated and difficult to perform with the increase of the number of qubits. Zhang et al~\cite{24} proposed a scheme for one-step implementation of an n-qubit controlled-phase gate in a superconducting quantum interference device system. The scheme focuses on that $n$ SQUID qubits simultaneously and nonidentically couple to a resonator mode and the microwave pulses, which requires individual addressing on each qubit. It means that $n$ classical microwave fields are needed to drive $n$ qubits in the same resonator, which poses a challenge to the present experimental condition as the number of qubits increases. Due to the large detuning, the gate speed is greatly limited to the order of $\mu s$. Moreover, the phase can not be tunable. It is just a $\pi$-phase gate. 

Here, we propose a scheme for realization of multiqubit tunable phase gate with only one step. This scheme differs remarkably from others due to the fact that we employ the quantum Zeno dynamics~\cite{29,30,31,32,33} and the distributed experimental setup where $n-1$ qubits located in $n-1$ different resonators respectively. Compared with previous proposals, our scheme owns several advantages as following: (i) individual addressing on each qudit is not required and only a classical microwave pulse is needed to drive the central qudit $A$, which greatly loosens the requirement for the experimental conditions; (ii) the time needed to complete the gate can be reached to the order of nanosecond, which is much faster than previous schemes~\cite{20,24}; (iii) in the non-resonant case, the phase is tunable. It can be adjusted by changing the Rabi frequency of the pulse applied to the target qubit, the detuning and the interaction time.

This paper is organized as follows. In Sec. II, we briefly introduce the our model with four-level quantum systems coupled to resonators where are connected by the common coupler, and how to realize the gate within such a system. In Sec. III, we give a brief discussion of the effectiveness of our model through numerical simulation. In Sec. IV, we generalize the model to $N$-qubits case. A concluding summary is presented in Sec. V.

\begin{figure}[t]
\begin{center}
\subfigure[]{\includegraphics[width=0.4\textwidth]{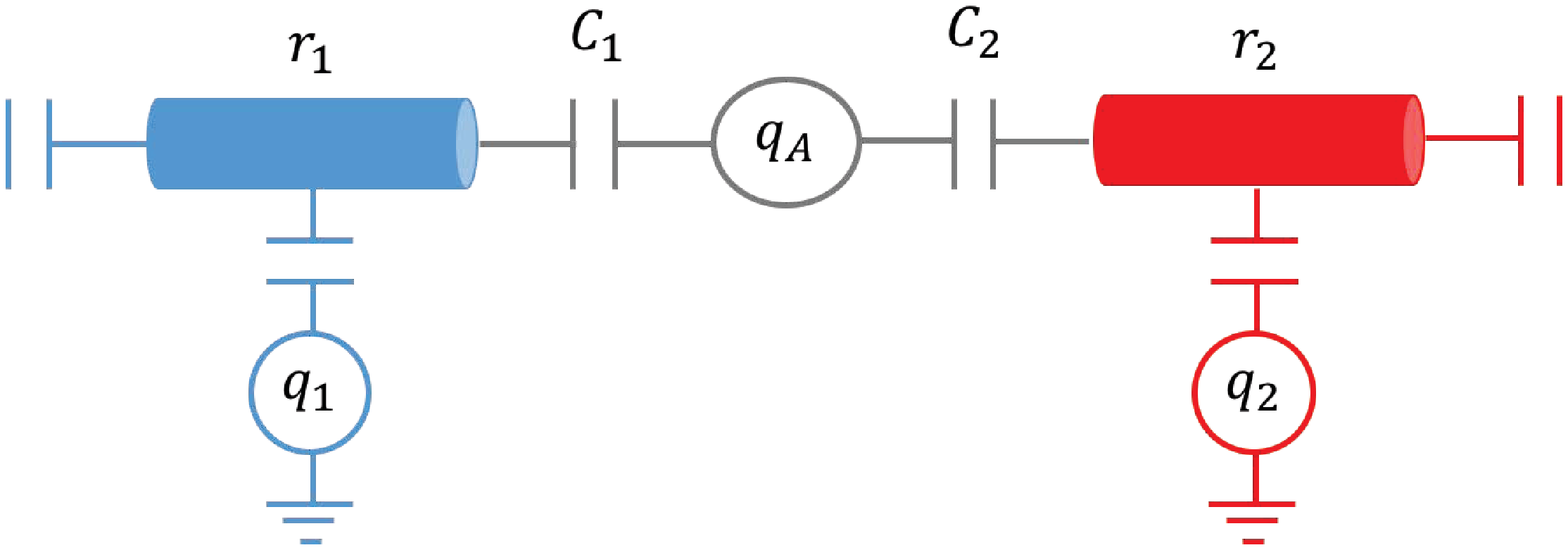}}
\subfigure[]{\includegraphics[width=0.4\textwidth]{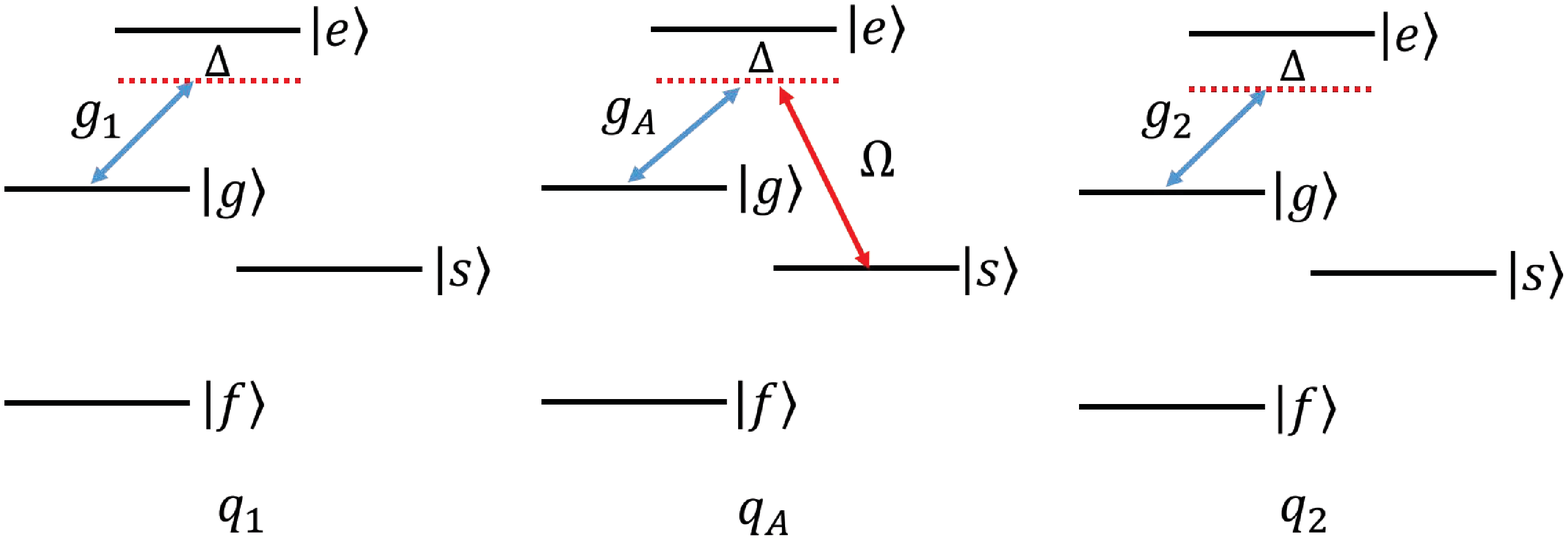}}
\subfigure[]{\includegraphics[width=0.18\textwidth]{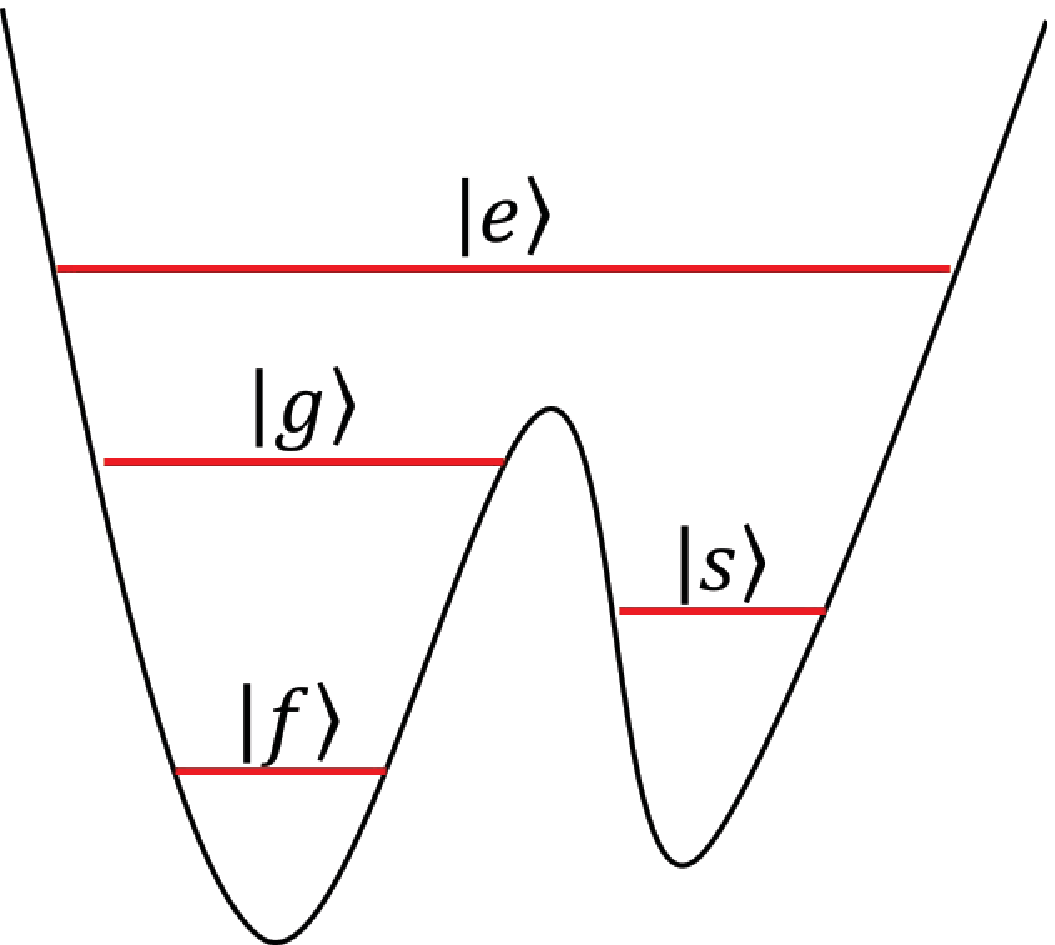}}
\end{center}
\caption{(a) Setup for the construction of our tunable phase gate. $q_1$ ($q_2$) is coupled to the high-quality resonator $r_1$ ($r_2$). The two resonators are interconnected by a qudit $q_A$ capacitively. $C_1, C_2$ represent capacitors. 
Here, $q_i$ ($i=1,2,A$) is fluxonium qudit ($d=4$). (b) The level configuration and relevant transitions of the qudits. The transition $|g_{A(1,2)}\rangle\longleftrightarrow|e_{A(1,2)}\rangle$ is coupled to the resonator mode with the coupling strength $g_{A(1,2)}$ and detuning $\Delta$. The transition $|s_A\rangle\longleftrightarrow|e_A\rangle$ is driven by a classical microwave pulse with Rabi frequency $\Omega$ and detuning $\Delta$. (c) The energy level diagram of a fluxonium qudit with the $\Lambda$-type four lowest levels $|f\rangle$, $|s\rangle$, $|g\rangle$ and $|e\rangle$.}
\label{fig:1}
\end{figure}

\section{model and effective dynamics}
We consider a system consisting of two resonators each hosting a fluxonium qudit~\cite{flux1,flux2,kim} and interconnected by a common fluxonium qudit $A$ capacitively, as shown in Fig.~\ref{fig:1}(a). The fluxonium qudit is biased properly to have four lowest levels, which are denoted by $|f_j\rangle$, $|s_j\rangle$, $|g_j\rangle$ and $|e_j\rangle$ ($j=1,2,A$), respectively [Fig.~\ref{fig:1}(b) and (c)]. For all the qudits, the resonator mode is off-resonant with the transition $|e\rangle\leftrightarrow|g\rangle$ while decoupled from the transition between any other two levels of fluxonium qudits. Here, $g_j$ ($j=1,2,A$) is the  coupling strength between the resonator mode and the $|e_j\rangle\leftrightarrow|g_j\rangle$ transition. $\Delta$ is the detuning between the $|e\rangle\leftrightarrow|g\rangle$ transition. For the qudit $A$, the transition $|e_A\rangle\leftrightarrow|s_A\rangle$ is driven dispersively by a classical microwave pulse with Rabi frequency $\Omega$ and detuning $\Delta$.
In the interaction picture, the Hamiltonian of the whole system can be written as 
\begin{equation}\label{eq1}
H=H_1+H_2,
\end{equation}
with
\begin{equation}
H_1=\Delta\sum_{i=1,2,A} |e\rangle_i \langle e|+\left(\Omega |s\rangle_A \langle e|+\mathrm{h.c.}\right),
\end{equation}
\begin{equation}
H_2=g_1a_1^\dag|g\rangle_1\langle e|+g_2a_2^\dag|g\rangle_2\langle e|+g_A\sum_{i=1,2}a_i^\dag|g\rangle_A\langle e|+\mathrm{h.c.},
\end{equation}
where $H$ is the total interaction of the whole system, $H_{1,2}$ is the interaction between the qudits and the resonators (or classical microwave pulse), and $a_j$ is the annihilation operator of resonator $j$. It is noted that $|f_i\rangle$ is decoupled with the qudit-resonator interaction, thus it disappears in Eq.(\ref{eq1}). 
For simplicity, we assume $g_j$ ($j=1,2,A$) and $\Omega$ are all real, and $g_1=g_2=g_A=g$.
To implement the three qubit quantum phase gate, we here use the asymmetric encoding scheme. The logic states of qubit 1 and qubit 2 are represented by the state $|f\rangle$ and $|g\rangle$, while the logic states of qubit $A$ are represented by $|f\rangle$ and $|s\rangle$. With this, the three qubit computational basis corresponds to $\{|f_1f_2f_A\rangle,|f_1g_2f_A\rangle,|f_1f_2s_A\rangle,|f_1g_2s_A\rangle,|g_1f_2f_A\rangle,|g_1g_2f_A\rangle,$ $|g_1f_2s_A\rangle,|g_1g_2s_A\rangle\}$.

First, we consider the case that the system is initially in the $|f_1f_2s_A\rangle|0,0\rangle_c$, where $|0,0\rangle_c$ denotes the vacuum state of the resonator mode 1 and 2 respectively. As a consequence, it will be constrained in the subspace spanned by $\{|f_1f_2s_A\rangle|0,0\rangle_c, |f_1f_2e_A\rangle|0,0\rangle_c, |f_1f_2g_A\rangle|1,0\rangle_c, |f_1f_2g_A\rangle|0,1\rangle_c\}$. In such a subspace, we can rewrite the Hamiltonian as
\begin{eqnarray}
H_1^\prime&=&\frac{\Delta}{2}(-|\phi_1\rangle+|\phi_2\rangle)(-\langle \phi_1|+\langle \phi_2|)\\
\nonumber&&+\left[\frac{\Omega}{\sqrt{2}}|f_1f_2s_A\rangle|0,0\rangle_c(-\langle \phi_1|+\langle \phi_2|)+\mathrm{h.c.}\right],
\end{eqnarray}
\begin{equation}
H_2^\prime=-\sqrt{2}g|\phi_1\rangle \langle\phi_1|+\sqrt{2}g|\phi_2\rangle \langle\phi_2|.
\end{equation}
Here,
\begin{equation}
|\phi_1\rangle=\frac{1}{2}\left(-\sqrt{2}|f_1f_2e_A\rangle|0,0\rangle_c+|f_1f_2g_A\rangle|1,0\rangle_c+ |f_1f_2g_A\rangle|0,1\rangle_c\right),
\end{equation}
\begin{equation}
|\phi_2\rangle=\frac{1}{2}\left(\sqrt{2}|f_1f_2e_A\rangle|0,0\rangle_c+|f_1f_2g_A\rangle|1,0\rangle_c+ |f_1f_2g_A\rangle|0,1\rangle_c\right),
\end{equation}
\begin{equation}
|\phi_3\rangle=\frac{1}{\sqrt{2}}\left(-|f_1f_2g_A\rangle|1,0\rangle_c+ |f_1f_2g_A\rangle|0,1\rangle_c\right),
\end{equation}
are the eigenstates of $H_2$ with the eigenvalues $-\sqrt{2}g$, $\sqrt{2}g$, $0$, respectively. Under the unitary transformation $e^{iH_2^\prime t}$, we further obtain
\begin{eqnarray}
H_1^{\prime\prime}&=&\frac{\Delta}{2}\left(|\phi_1\rangle \langle\phi_1|+|\phi_2\rangle \langle\phi_2|\right.\\\nonumber
&&\left.-|\phi_1\rangle \langle\phi_2|e^{-2\sqrt{2}igt}-|\phi_2\rangle \langle\phi_1|e^{2\sqrt{2}igt}\right)\\\nonumber
&&+\left[\frac{\Omega}{\sqrt{2}}|f_1f_2s_A\rangle|0,0\rangle_c(-\langle \phi_1|e^{\sqrt{2}igt}+\langle \phi_2|e^{-\sqrt{2}igt})+\mathrm{h.c.}\right].
\end{eqnarray}
Assuming the conditions $g\gg\Omega$ are satisfied, we can readily discard the fast-oscillating terms in $H_1^{\prime\prime}$, then obtain the effective Hamiltonian
\begin{equation}
H_{1,\mathrm{eff}}^{\prime\prime}=\frac{\Delta}{2}\left(|\phi_1\rangle \langle\phi_1|+|\phi_2\rangle \langle\phi_2|\right).
\end{equation}
This effective Hamiltonian does nothing to the initial state $|f_1f_2s_A\rangle|0,0\rangle_c$, thus the initial state remains unchanged.

Next, we consider the case that the system is initially in the state $|f_1g_2s_A\rangle |0,0\rangle_c$. The system will evolve in the subspace $\{|f_1g_2s_A\rangle|0,0\rangle_c, |f_1g_2e_A\rangle|0,0\rangle_c, |f_1g_2g_A\rangle|1,0\rangle_c, |f_1g_2g_A\rangle|0,1\rangle_c,$ $|f_1e_2g_A\rangle|0,0\rangle_c\}$. The relevant Hamiltonian of the system can be rewritten as
\begin{eqnarray}
\bar{H}_1&=&\Delta\left[N_+\left(|\phi_1^\prime\rangle+|\phi_2^\prime\rangle\right)-N_-\left(|\phi_3^\prime\rangle+|\phi_4^\prime\rangle\right)\right]\\\nonumber
&&\times\left[N_+\left(\langle\phi_1^\prime|+\langle\phi_2^\prime|\right)-N_-\left(\langle\phi_3^\prime|+\langle\phi_4^\prime|\right)\right]\\\nonumber
&&+\Delta\left[-N_+^\prime\left(|\phi_1^\prime\rangle+|\phi_2^\prime\rangle\right)+N_-^\prime\left(|\phi_3^\prime\rangle+|\phi_4^\prime\rangle\right)\right]\\\nonumber
&&\times\left[-N_+^\prime\left(\langle\phi_1^\prime|+\langle\phi_2^\prime|\right)+N_-^\prime\left(\langle\phi_3^\prime|+\langle\phi_4^\prime|\right)\right]
\\\nonumber
&&+\left(\Omega|f_1g_2s_A\rangle|0,0\rangle_c\left[N_+\left(\langle\phi_1^\prime|+\langle\phi_2^\prime|\right)\right.\right.\\\nonumber
&&\left.\left.-N_-\left(\langle\phi_3^\prime|+\langle\phi_4^\prime|\right)\right]+\mathrm{h.c.}\right),
\end{eqnarray}
\begin{equation}
\bar{H}_2=\sum_{i=1}^4\lambda_i|\phi_i^\prime\rangle \langle \phi_i^\prime|,
\end{equation}
where $N_+=\frac{\sqrt{5+\sqrt{5}}}{2\sqrt{5}}$, $N_-=\frac{\sqrt{5-\sqrt{5}}}{2\sqrt{5}}$, $N_+^\prime=\frac{(1-\sqrt{5})\sqrt{5+\sqrt{5}}}{4\sqrt{5}}$, $N_-^\prime=\frac{(1+\sqrt{5})\sqrt{5-\sqrt{5}}}{4\sqrt{5}}$.
The eigenvectors of the interaction Hamiltonian $H_2$ are listed as following
\begin{eqnarray}
\nonumber
|\phi_1^\prime\rangle&=&\frac{1}{\sqrt{5+\sqrt{5}}}\left(\frac{1+\sqrt{5}}{2}|f_1g_2e_A\rangle|0,0\rangle_c-|f_1g_2g_A\rangle|1,0\rangle_c\right.\\
&&\left.-\frac{1+\sqrt{5}}{2}|f_1g_2g_A\rangle|0,1\rangle_c+|f_1e_2g_A\rangle|0,0\rangle_c\right),
\end{eqnarray}
\begin{eqnarray}
\nonumber
|\phi_2^\prime\rangle&=&\frac{1}{\sqrt{5+\sqrt{5}}}\left(\frac{1+\sqrt{5}}{2}|f_1g_2e_A\rangle|0,0\rangle_c+|f_1g_2g_A\rangle|1,0\rangle_c\right.\\
&&\left.+\frac{1+\sqrt{5}}{2}|f_1g_2g_A\rangle|0,1\rangle_c+|f_1e_2g_A\rangle|0,0\rangle_c\right),
\end{eqnarray}
\begin{eqnarray}
\nonumber
|\phi_3^\prime\rangle&=&\frac{1}{\sqrt{5-\sqrt{5}}}\left(\frac{1-\sqrt{5}}{2}|f_1g_2e_A\rangle|0,0\rangle_c+|f_1g_2g_A\rangle|1,0\rangle_c\right.\\
&&\left.+\frac{1-\sqrt{5}}{2}|f_1g_2g_A\rangle|0,1\rangle_c+|f_1e_2g_A\rangle|0,0\rangle_c\right),
\end{eqnarray}
\begin{eqnarray}
\nonumber
|\phi_4^\prime\rangle&=&\frac{1}{\sqrt{5-\sqrt{5}}}\left(\frac{1-\sqrt{5}}{2}|f_1g_2e_A\rangle|0,0\rangle_c-|f_1g_2g_A\rangle|1,0\rangle_c\right.\\
&&\left.-\frac{1-\sqrt{5}}{2}|f_1g_2g_A\rangle|0,1\rangle_c+|f_1e_2g_A\rangle|0,0\rangle_c\right),
\end{eqnarray}
with eigenvalues $\lambda_1=-\frac{1+\sqrt{5}}{2}g$, $\lambda_2=\frac{1+\sqrt{5}}{2}g$, $\lambda_3=\frac{1-\sqrt{5}}{2}g$, $\lambda_4=-\frac{1-\sqrt{5}}{2}g$. Similarly, under the unitary transformation $e^{i\bar{H}_2t}$ and the condition $g\gg\Omega$, the $\bar{H}_1$ becomes
\begin{eqnarray}
\nonumber
\bar{H}_{1,\mathrm{eff}}&=&\Delta (N_+^2+N_+^{\prime 2})(|\phi_1^\prime\rangle \langle \phi_1^\prime|+|\phi_2^\prime\rangle \langle \phi_2^\prime|)\\
&&+\Delta (N_-^2+N_-^{\prime 2})(|\phi_3^\prime\rangle \langle \phi_3^\prime|+|\phi_4^\prime\rangle \langle \phi_4^\prime|).
\end{eqnarray}
Obviously, the effective Hamiltonian also does nothing to the initial state $|f_1g_2s_A\rangle|0,0\rangle_c$ and the $|f_1g_2s_A\rangle|0,0\rangle_c$ do not undergo any change during the interaction. Moreover, it is noted that the system will undergo the similar evolution with the initial state $|g_1f_2s_A\rangle|0,0\rangle_c$ due to the exchange symmetry between qubit 1 and 2.

Furthermore, if the system is assumed to be prepared in the state $|g_1g_2s_A\rangle|0,0\rangle_c$, the system will be constrained in the subspace spanned by 
$\{|g_1g_2s_A\rangle|0,0\rangle_c, |g_1g_2e_A\rangle|0,0\rangle_c, |g_1g_2g_A\rangle|1,0\rangle_c, |e_1g_2g_A\rangle|0,0\rangle_c$, $|g_1g_2g_A\rangle|0,1\rangle_c$, $|g_1e_2g_A\rangle|0,0\rangle_c\}$. The Hamiltonian of this subsystem is dominated by
\begin{eqnarray}
\tilde{H}_1&=&\frac{\Delta}{3}(|\phi_1^{\prime\prime}\rangle+|\phi_2^{\prime\prime}\rangle-|\phi_5^{\prime\prime}\rangle)(\langle\phi_1^{\prime\prime}|+\langle\phi_2^{\prime\prime}|-\langle\phi_5^{\prime\prime}|)\\\nonumber
&&+\frac{\Delta}{2}\left[\frac{1}{3}(|\phi_1^{\prime\prime}\rangle+|\phi_2^{\prime\prime}\rangle+2|\phi_5^{\prime\prime}\rangle)(\langle\phi_1^{\prime\prime}|+\langle\phi_2^{\prime\prime}|+2\langle\phi_5^{\prime\prime}|)\right.\\\nonumber
&&\left.+(|\phi_3^{\prime\prime}\rangle+|\phi_4^{\prime\prime}\rangle)(\langle\phi_3^{\prime\prime}|+\langle\phi_4^{\prime\prime}|)\right]\\\nonumber
&&+\left[\frac{\Omega}{\sqrt{3}}|g_1g_2s_A\rangle|0,0\rangle_c(\langle\phi_1^{\prime\prime}|+\langle\phi_2^{\prime\prime}|-\langle\phi_5^{\prime\prime}|)+\mathrm{h.c.}\right],
\end{eqnarray}
\begin{equation}
\tilde{H}_2=\sum_{i=1}^5\lambda_i^
\prime|\phi_i^{\prime\prime}\rangle \langle \phi_i^{\prime\prime}|,
\end{equation}
where the corresponding eigenvectors of $H_2$ in such a subsystem are
\begin{eqnarray}
\nonumber
|\phi_1^{\prime\prime}\rangle&=&\frac{1}{2\sqrt{3}}\left(2|g_1g_2e_A\rangle|0,0\rangle_c-\sqrt{3}|g_1g_2g_A\rangle|1,0\rangle_c\right.\\\nonumber
&&+|e_1g_2g_A\rangle|0,0\rangle_c-\sqrt{3}|g_1g_2g_A\rangle|0,1\rangle_c\\
&&\left.+|g_1e_2g_A\rangle|0,0\rangle_c\right),
\end{eqnarray}
\begin{eqnarray}
\nonumber
|\phi_2^{\prime\prime}\rangle&=&\frac{1}{2\sqrt{3}}\left(2|g_1g_2e_A\rangle|0,0\rangle_c+\sqrt{3}|g_1g_2g_A\rangle|1,0\rangle_c\right.\\\nonumber
&&+|e_1g_2g_A\rangle|0,0\rangle_c+\sqrt{3}|g_1g_2g_A\rangle|0,1\rangle_c\\
&&\left.+|g_1e_2g_A\rangle|0,0\rangle_c\right),
\end{eqnarray}
\begin{eqnarray}
\nonumber
|\phi_3^{\prime\prime}\rangle&=&\frac{1}{2}(|g_1g_2g_A\rangle|1,0\rangle_c-|e_1g_2g_A\rangle|0,0\rangle_c\\
&&-|g_1g_2g_A\rangle|0,1\rangle_c+|g_1e_2g_A\rangle|0,0\rangle_c),
\end{eqnarray}
\begin{eqnarray}
\nonumber
|\phi_4^{\prime\prime}\rangle&=&\frac{1}{2}(-|g_1g_2g_A\rangle|1,0\rangle_c-|e_1g_2g_A\rangle|0,0\rangle_c\\
&&+|g_1g_2g_A\rangle|0,1\rangle_c+|g_1e_2g_A\rangle|0,0\rangle_c),
\end{eqnarray}
\begin{equation}
|\phi_5^{\prime\prime}\rangle=\frac{1}{\sqrt{3}}(-|g_1g_2e_A\rangle|0,0\rangle_c+|e_1g_2g_A\rangle|0,0\rangle_c+|g_1e_2g_A\rangle|0,0\rangle_c),
\end{equation}
with eigenvalues $\lambda_1^\prime=-\sqrt{3}g, \lambda_2^\prime=\sqrt{3}g, \lambda_3^\prime=-g, \lambda_4^\prime=g, \lambda_5^\prime=0$.
In the interaction picture with respect to $\tilde{H}_2$, considering the condition $g\gg\Omega$ and discarding the fast-oscillating terms, then we can obtain
\begin{equation}
\tilde{H}_1^\prime=\Delta|\phi_5^{\prime\prime}\rangle \langle \phi_5^{\prime\prime}|-\left(\frac{\Omega}{\sqrt{3}}|g_1g_2s_A\rangle|0,0\rangle_c\langle\phi_5^{\prime\prime}|+\mathrm{h.c.}\right).
\end{equation}

\emph{Non-resonant case:}
Set $\Delta\gg\Omega$, then there are no any energy exchange between the state $|g_1g_2s_A\rangle|0,0\rangle_c$ and $|\phi_5^{\prime\prime}\rangle$ due to the large detuning. Consequently, the effective Hamiltonian of the subsystem
\begin{equation}
\tilde{H}_{1,\mathrm{eff}}^\prime=\frac{\Omega^2}{3\Delta}|g_1g_2s_A\rangle|0,0\rangle_c \langle 0,0|\langle g_1g_2s_A|
\end{equation}
is obtained. Under the action of $\tilde{H}_{1,\mathrm{eff}}^\prime$, we obtain $|g_1g_2s_A\rangle|0,0\rangle_c\rightarrow\exp(i\Omega^2t/{3\Delta})|g_1g_2s_A\rangle|0,0\rangle_c$. The other computational states $|f_1x_2f_A\rangle$ and $|g_1x_2f_A\rangle$ (where $x=f, g$) are decoupled from the Hamiltonian and do not undergo any change during the evolution of the system. In this way, the system keeps in the initial state with an tunable additional phase shift. Therefore, we obtain a three qubit tunable phase gate
\begin{eqnarray}
|f_1f_2f_A\rangle\rightarrow|f_1f_2f_A\rangle\nonumber\\
|f_1g_2f_A\rangle\rightarrow|f_1g_2f_A\rangle\nonumber\\
|f_1f_2s_A\rangle\rightarrow|f_1f_2s_A\rangle\nonumber\\
|f_1g_2s_A\rangle\rightarrow|f_1f_2s_A\rangle\nonumber\\
|g_1f_2f_A\rangle\rightarrow|g_1f_2f_A\rangle\nonumber\\
|g_1g_2f_A\rangle\rightarrow|g_1g_2f_A\rangle\nonumber\\
|g_1f_2s_A\rangle\rightarrow|g_1f_2s_A\rangle\nonumber\\
|g_1g_2s_A\rangle\rightarrow e^{i\delta}|g_1g_2s_A\rangle
\end{eqnarray}
with $\delta=\Omega^2t/{3\Delta}$ being the phase. Additionally, if $\delta=\pi$, this transformation plus the Hadamard gate on the qubit $A$ with $|f_A\rangle\rightarrow(|f_A\rangle+|s_A\rangle)/\sqrt{2}$, $|s_A\rangle\rightarrow(|f_A\rangle-|s_A\rangle)/\sqrt{2}$, we can obtain a three qubit Toffoli gate.

\emph{Resonant case:}
When $\Delta=0$, after time $t$, the state of the system becomes
$\cos{(\Omega t/\sqrt{3})}|g_1g_2s_A\rangle|0,0\rangle_c+i\sin{(\Omega t/\sqrt{3})}|\phi_5^{\prime\prime}\rangle$. After a full cycle of Rabi oscillation, i.e., $t=\sqrt{3}\pi/\Omega$, we have $-|g_1g_2s_A\rangle|0,0\rangle_c$. Thus, the system returns to the initial state with an additional phase shift $\pi$. In this way, we obtain a three-qubit controlled phase gate
\begin{equation}
U_p=e^{i\pi |g_1g_2s_A\rangle|0,0\rangle_c\langle 0,0|\langle g_1g_2s_A|},
\end{equation}
in which, if and only if the three qubits are in the state $|g_1g_2s_A\rangle|0,0\rangle_c$, the system undergoes a phase shift $\pi$.
\begin{figure}[t]
\begin{center}
\subfigure[]{\includegraphics[width=0.22\textwidth]{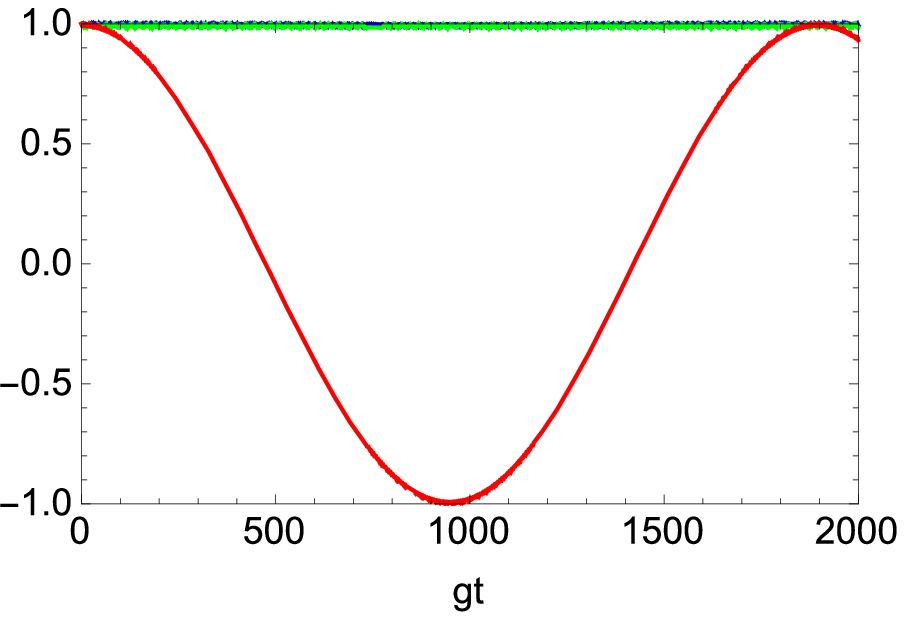}}
\subfigure[]{\includegraphics[width=0.22\textwidth]{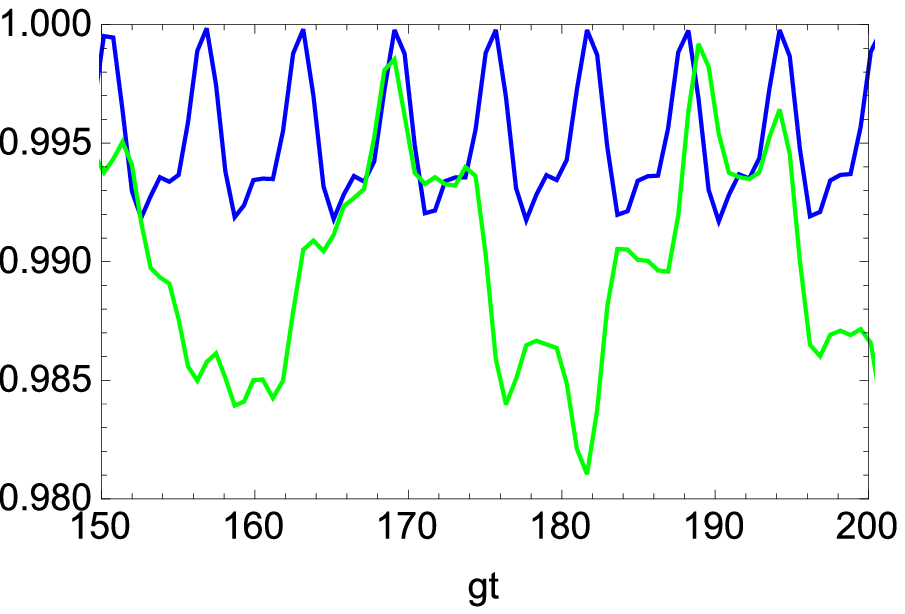}}
\subfigure[]{\includegraphics[width=0.22\textwidth]{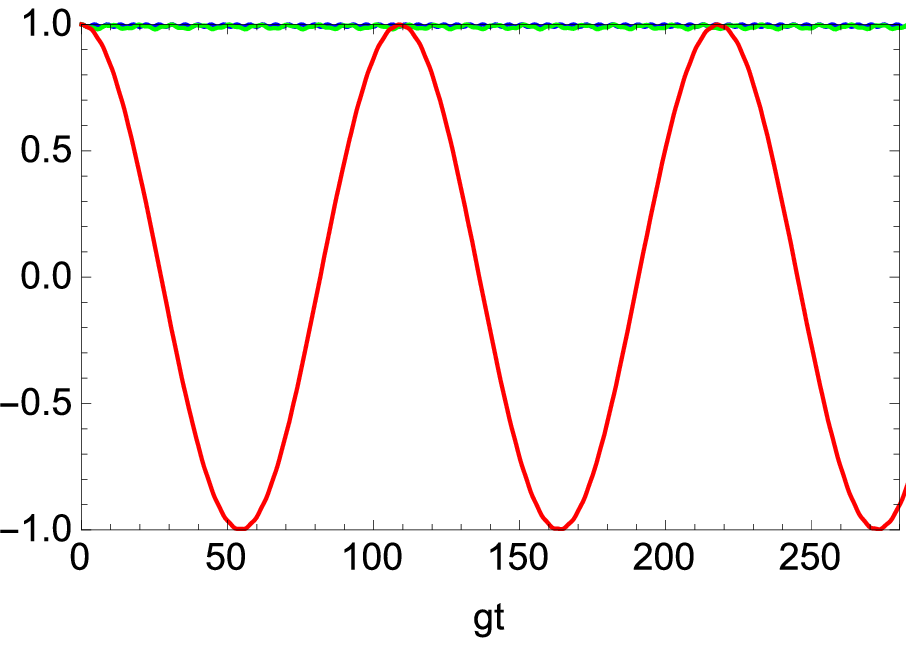}}
\subfigure[]{\includegraphics[width=0.22\textwidth]{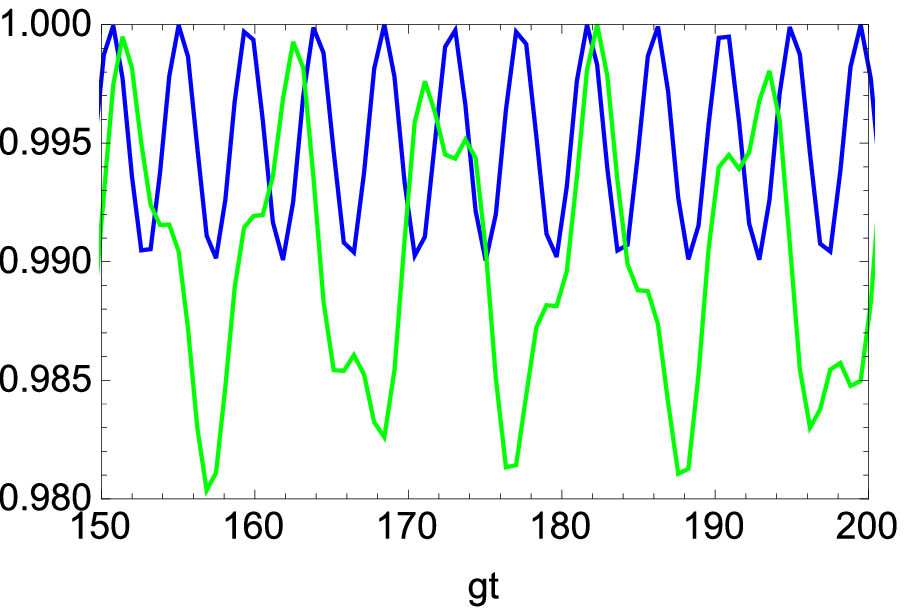}}
\end{center}
\caption{Real parts of the coefficients of the state $|f_1f_2s_A\rangle|0,0\rangle_c$ (blue line), $|f_1g_2s_A\rangle|0,0\rangle_c$ (green line), $|g_1g_2s_A\rangle|0,0\rangle_c$ (red line) versus the evolution time under (a) non-resonant condition; (c) resonant condition. (b) and (d) depicts the enlarged part from (a) and (c) respectively. The parameters are typically set as: $\Omega=0.1g$, $\Delta=g$ (non-resonant case).}
\label{fig:2}
\end{figure}

\section{discussions and numerical analysis}
In order to validate the feasibility of the above theoretical analysis, we perform a direct numerical simulation of the Schr\"{o}dinger equation with the original Hamiltonian Eq.~(\ref{eq1}) (without decoherence). In non-resonant case, we choose the typical parameters: $\Omega=0.1g$ and $\Delta=g$. In the simulation, we calculated the temporal evolutions of the system beginning with three distinct initial states $|f_1f_2s_A\rangle|0,0\rangle_c$, $|f_1g_2s_A\rangle|0,0\rangle_c$ and $|g_1g_2s_A\rangle|0,0\rangle_c$. As shown in the Fig.\ref{fig:2}(a), the blue, green and red lines represent the real parts of the coefficients of the basis states $|f_1f_2s_A\rangle|0,0\rangle_c$, $|f_1g_2s_A\rangle|0,0\rangle_c$ and $|g_1g_2s_A\rangle|0,0\rangle_c$, respectively. It is seen that, the system returns to its initial state but obtains a global phase shift $\pi$ at the time $\tau=3\pi\Delta/\Omega^2=300\pi/g$, when the system is initially prepared in the state $|g_1g_2s_A\rangle|0,0\rangle_c$, while it is almost unchanged for the initial state $|f_1f_2s_A\rangle|0,0\rangle_c$ and $|f_1g_2s_A\rangle|0,0\rangle_c$. Furthermore, we also consider the resonant case with parameters $\Omega=0.1g$ and $\Delta=0$ in Fig.\ref{fig:2}(c). At the time $\tau^\prime=\sqrt{3}\pi/\Omega=10\sqrt{3}\pi/g$, the system returns to the initial state with an additional phase $\pi$, which is much shorter than the time required in the non-resonant case. In particular, Fig.\ref{fig:2}(b) and (d) shows the enlarged part of Fig.\ref{fig:2}(a) and (c) respectively. It represents that state $|f_1f_2s_A\rangle|0,0\rangle_c$ and $|f_1g_2s_A\rangle|0,0\rangle_c$ are  almost unchanged during the process. 

\begin{figure}[t]
\begin{center}
\subfigure[]{\includegraphics[width=0.33\textwidth]{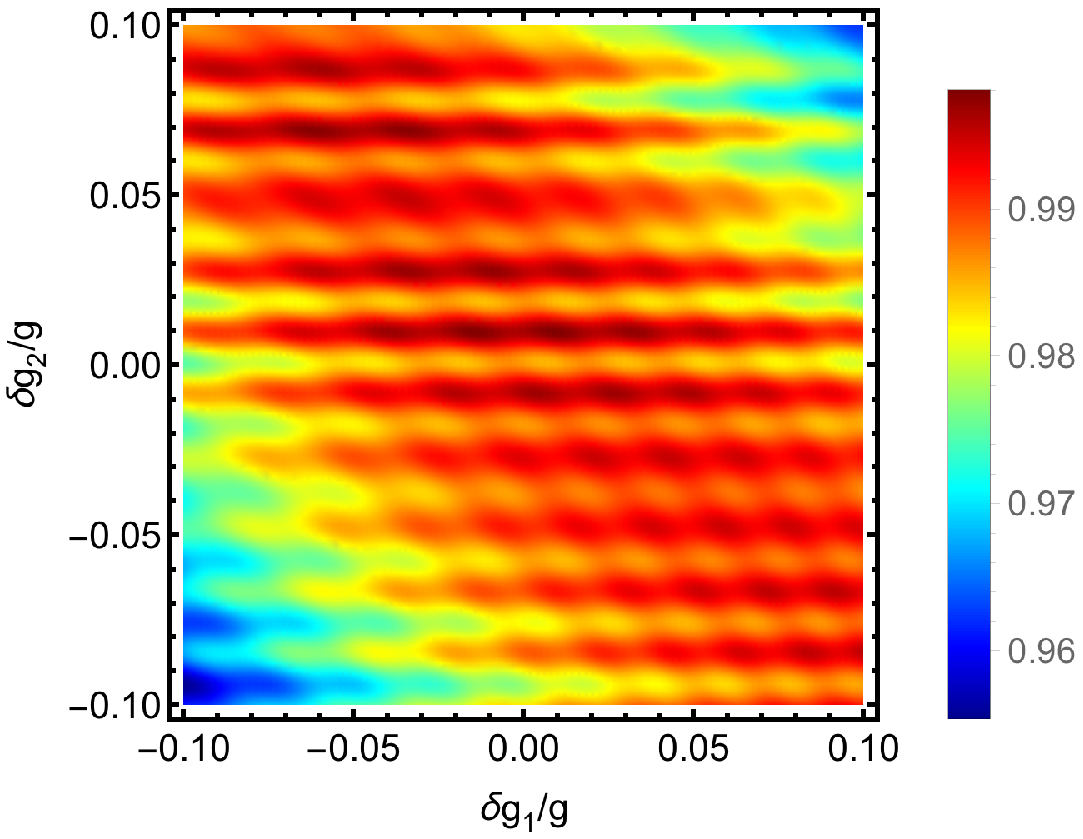}}
\subfigure[]{\includegraphics[width=0.33\textwidth]{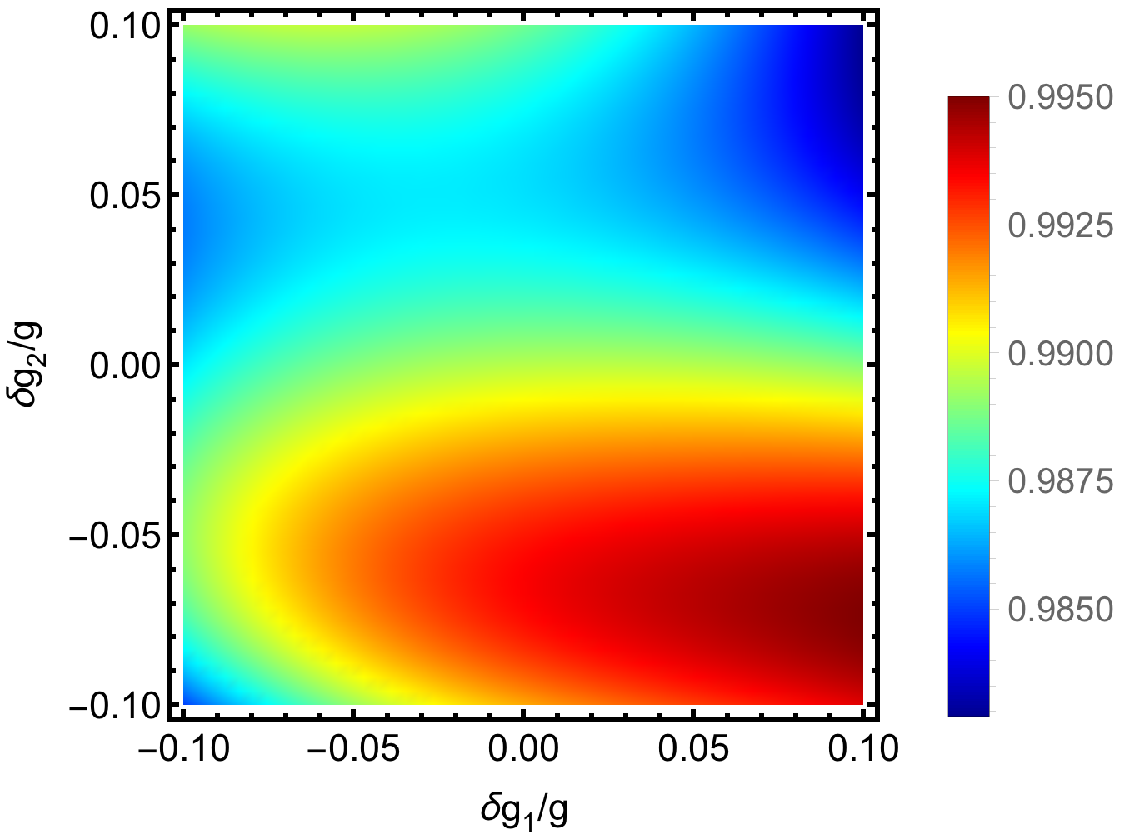}}
\end{center}
\caption{The fidelity of the three-qubit phase gate vs parameter (coupling constants $g_1$, $g_2$) errors under (a) non-resonant condition; (b) resonant condition.}
\label{fig:3}
\end{figure}

The validity of our scheme is based on the assumption that all the coupling strengths of qudit-resonator mode are equal, namely, $g_1=g_2=g_A=g$. However, there could be deviation in the parameters in a practical situation. These errors result in the mismatch of the coupling constants $g_1$ and $g_2$. So we should consider the influence of the
 deviation from theoretical situation $g$ on the fidelity of the three-qubit phase gate, which is defined as $F=|\langle\psi(\tau)|U_p|\Psi(0)\rangle|^2$, where $|\Psi(0)\rangle$ is the initial state of the qubits and $|\psi(\tau)\rangle$ is the final state under the evolution of the original Hamiltonian Eq.(\ref{eq1}) at time $\tau$. Here we consider a general input state
 \begin{eqnarray}
|\Psi(0)\rangle=c_1|f_1f_2f_A\rangle+c_2|f_1g_2f_A\rangle+c_3|f_1f_2s_A\rangle+c_4|f_1g_2s_A\rangle\\\nonumber
+c_5|g_1f_2f_A\rangle+c_6|g_1g_2f_A\rangle+c_7|g_1f_2s_A\rangle+c_8|g_1g_2s_A\rangle,
\end{eqnarray}
where $c_i$ is the corresponding amplitude of probability obeying the normalization $\sum_i |c_i|^2=1$. Without loss of generality, we select $c_1=\frac{1}{6}, c_2=\frac{\sqrt{2}}{6}, c_3=\frac{\sqrt{3}}{6}, c_4=\frac{1}{3}, c_5=\frac{\sqrt{5}}{6}, c_6=\frac{\sqrt{6}}{6}, c_7=\frac{\sqrt{7}}{6}, c_8=\frac{\sqrt{2}}{3}$
  for the present simulation.  Figure \ref{fig:3} shows how the deviation of the parameter influence the fidelity of the phase gate within the non-resonant (Fig.\ref{fig:3}a) and resonant case (Fig.\ref{fig:3}b). A deviation $|\delta g_{1(2)}|=10\%g$ only causes a reduction smaller than 5$\%$ in the fidelity. It is apparent that the fidelity of phase gate is always higher than 95$\%$ under various deviations of the selected parameters. Thus our scheme is very robust against some errors which occurred in a practical case. 

\begin{figure}[t]
\begin{center}
\subfigure[]{\includegraphics[width=0.36\textwidth]{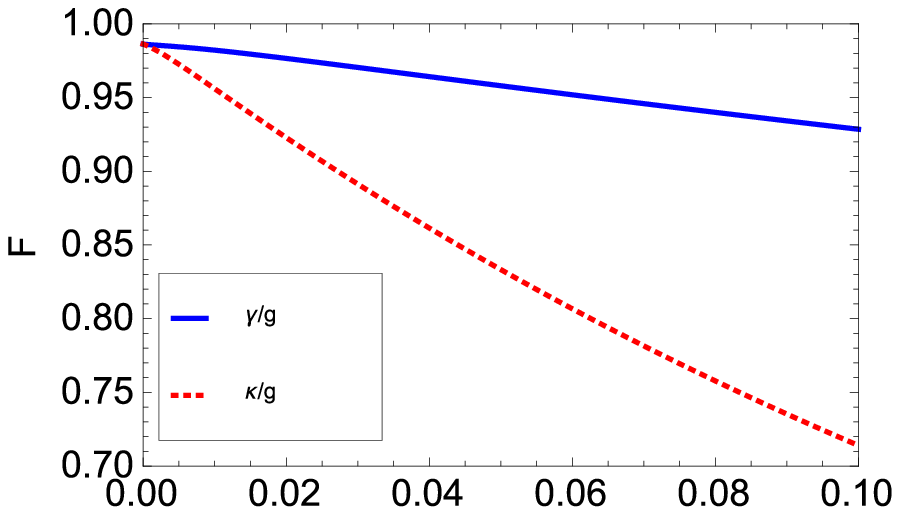}}
\subfigure[]{\includegraphics[width=0.36\textwidth]{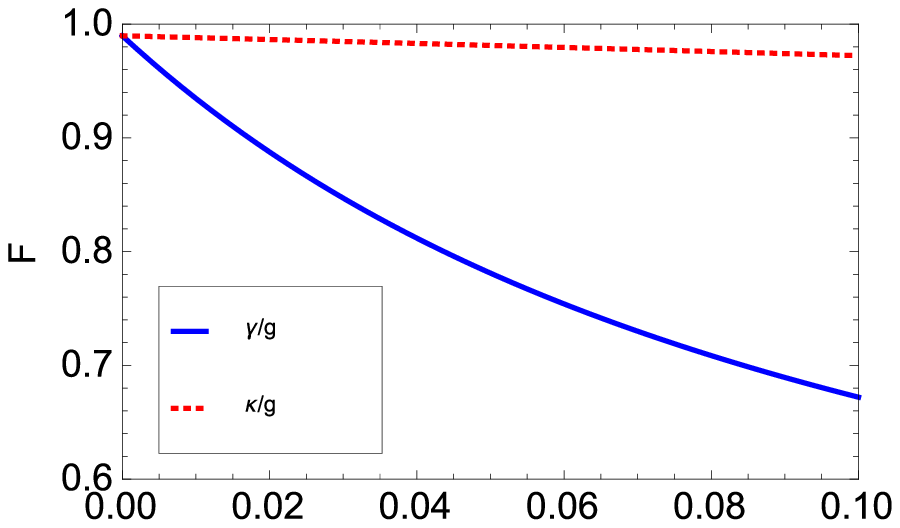}}
\end{center}
\caption{The influence of various decoherence process such as the decay of the resonator modes and the relaxation of the qudits on the fidelity of the three-qubit phase gate under (a) non-resonant condition; (b) resonant condition.}
\label{fig:4}
\end{figure}

Until now, we only consider the ideal case and various decoherence effects are not involved in the above discussions. The decoherence is induced by the decay of the cavities, and the relaxation of the qubits. Taking the decoherence into account, the whole system is determined by the master equation
\begin{eqnarray}
\dot{\rho} = -i[H,\rho]+\sum_{i=1,2}^2\kappa_iL[\hat{a}_i]+\sum_{i=1,2,A}\sum_{n=g,s,f}\gamma_{n,i}L[\sigma^-_{n,i}],
\end{eqnarray}
where $L[\hat{a}_i]=a_i\rho a_i^\dag-a_i^\dag a_i\rho/2-\rho a_i^\dag a_i/2$, $L[\sigma^-_{n,i}]=\sigma^-_{n,i}\rho\sigma^+_{n,i}-\sigma^+_{n,i}\sigma^-_{n,i}\rho/2-\rho\sigma^+_{n,i}\sigma^-_{n,i}/2$, $\sigma^-_{n,i}=|n_i\rangle\langle e_i|$. $\kappa_i$ is the photon decay rate of the $i$th cavity, $\gamma_{n,i}$ is the energy relaxation rate of the $j$th qubit for the decay path $|e\rangle\rightarrow|n\rangle$. We assume $\kappa_i=\kappa$ and $\gamma_{n,i}=\gamma$ for simplicity. The fidelity of the three-qubit controlled-phase gate implemented in the presence of the decoherence
can be defined as
\begin{equation}
F=\langle \Psi(0)|U_p^\dag \rho^\prime(t=\tau)U_p|\Psi(0)\rangle,
\end{equation}
where $\rho^\prime(t)$ represents the temporal reduced density matrix (obtained by tracing out the cavity mode part). In Fig.\ref{fig:4}, we plot the fidelity $F$ versus the decays $\kappa$ and $\gamma$. We can see that the fidelity is still larger than $70\%$ for $\kappa=\gamma=0.1g$. In the non-resonant situation, the energy relaxation of the qubits is greatly suppressed due to the large detuning (Fig.\ref{fig:4}a). The subspaces involved during the whole process include the excited states of the cavities, which greatly influence the fidelity of the phase gate. However, in the resonant case, the results are reverse. The energy relaxation of the qubits becomes the main decoherence source due to the resonant interaction, as shown in Fig.\ref{fig:4}b. In real circuit QED system, strong coupling between superconducting qubit and resonator  can be achieved with $g=2\pi\times360$MHz~\cite{18}, and $\kappa^{-1}=1\mu s$, $\gamma^{-1}=25\mu s$~\cite{34,35}. With these parameters, it seems that the present scheme with a high fidelity larger than 95\% could be feasible in an experiment. Furthermore, in the resonant case, the $\pi$ phase gate can be realized only in 24 $ns$, which is much shorter than the time needed in previous scheme~\cite{20,24}.

\section{generalization to $N$-qubit phase gate}
\begin{figure}[t]
\centering
\includegraphics[width=0.35\textwidth]{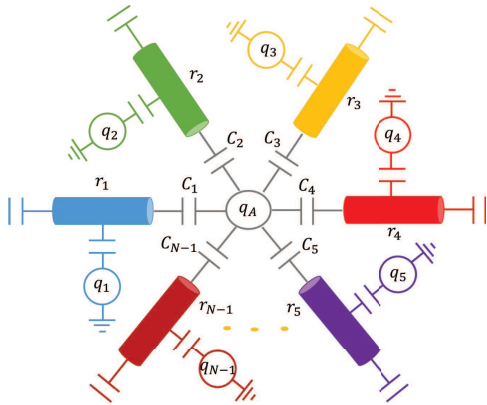}
\caption{The potential experimental setup of the $N$-qubit tunable phase gate. $N-1$ resonators each hosting a qudit are coupled to a common qudit $A$ capacitively.}
\label{fig:5}
\end{figure}

\begin{figure}[t]
\begin{center}
\subfigure[]{\includegraphics[width=0.35\textwidth]{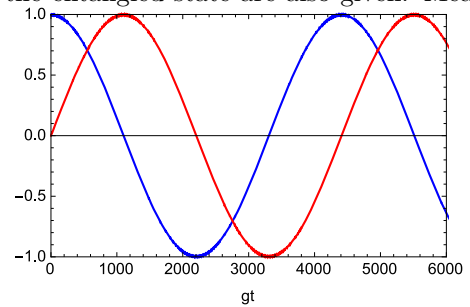}}
\subfigure[]{\includegraphics[width=0.34\textwidth]{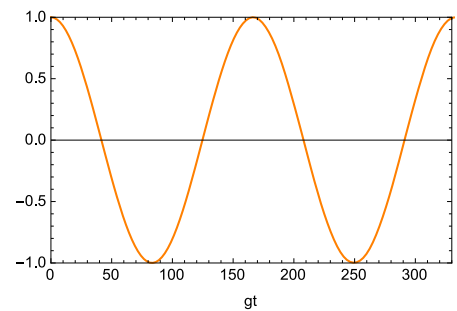}}
\end{center}
\caption{(a) In the non-resonant case,  the real part (blue line) and imaginary part (red line) of the coefficients of the state $|g_1g_2g_3g_4g_5g_6s_A\rangle|0,0,0,0,0,0\rangle_c$ versus the evolution time. (b) In the resonant case, the real part (orange line) of the coefficient of the state $|g_1g_2g_3g_4g_5g_6s_A\rangle|0,0,0,0,0,0\rangle_c$ versus the evolution time.}
\label{fig:6}
\end{figure}

We note that the scheme can be generalized to realize $N$-qubit phase gate. The potential experiment configurations are depicted in Fig.\ref{fig:5}. We assume that $N-1$ resonators each hosting a qudit are coupled to a common qudit $A$ capacitively. The level configuration of all qudits is the same as the case in Fig.\ref{fig:1}(b). Then the Hamiltonian reads
\begin{equation}\label{eq30}
H_N=H_N^1+H_N^2,
\end{equation}
with
\begin{equation}
H_N^1=\Delta\sum_{i=1}^{N-1} |e\rangle_i \langle e|+\Delta|e\rangle_A\langle e|+\left(\Omega |s\rangle_A \langle e|+\mathrm{h.c.}\right),
\end{equation}
\begin{equation}
H_N^2=\sum_{i}^{N-1}g_ia_i^\dag|g\rangle_i\langle e|+g_A\sum_{i=1}^{N-1}a_i^\dag|g\rangle_A\langle e|+\mathrm{h.c.},
\end{equation}
where $a_i$($i=1,2,3,N-1$) is the annihilation operator for photons in the resonator $i$ and $g_i$ is the coupling constant for qudit $i$ associated with the corresponding quantized resonator modes. Without loss of generality, we choose $g_i=g_A=g$ in the following calculation. To implement the $N$-qubit quantum phase gate, we here use the asymmetric encoding scheme. The logic states of qubit $i$($i=1,2,3,...,N-1$) are represented by the state $|f\rangle$ and $|g\rangle$, while the logic states of qubit $A$ are represented by $|f\rangle$ and $|s\rangle$. Under the condition $g,\Delta\gg\Omega$, taking the similar procedure above, we can find that if and only if the $N$ qubits are in the state $|g_1g_2g_3...g_{N-1}s_A\rangle$, the system undergoes a phase shift $\exp(i\Omega^2t/N\Delta)$. Especially, in the resonant case, implementing the $N$-qubit $\pi$-phase gate requires time $gt=10\sqrt{N}\pi$.
In order to test the effectiveness of our proposal, we consider specifically, for example, the case of 7 qubit. In Fig.\ref{fig:6}, we plot the time-evolution behaviors of the real part (blue line) and imaginary part (red line) of the state $|g_1g_2g_3g_4g_5g_6s_A\rangle|0,0,0,0,0,0\rangle_c$ under the evolution of the total Hamiltonian Eq.(\ref{eq30}). From figure.\ref{fig:6}a, it is easily seen that at scaled time $gt\approx2200$ the state $|g_1g_2g_3g_4g_5g_6s_A\rangle$ acquires a $\pi$-phase shift, which agrees with our theoretical value $700\pi$ very well. In figure.\ref{fig:6}b, we plot the time-evolution behaviors of the real part (orange line) of $|g_1g_2g_3g_4g_5g_6s_A\rangle|0,0,0,0,0,0\rangle_c$ within resonant case. The time needed to complete the 7-qubit phase gate only requires 36 $ns$. The results match with the theoretical value very well. Therefore, our effective model is valid.

\section{summary}
In summary, we have proposed a scheme for implementation of the multiqubit tunable phase gate in a circuit QED setup where two resonators each hosting a qudit are coupled to a common qudit. Taking advantage of quantum Zeno dynamics and asymmetric encoding the logic state, the multiqubit tunable phase gate can be completed in only one step without individual addressing on each qudit during the whole process. Only a classical microwave pulse is needed, no matter how many qudits are involved. We have considered the our model under the non-resonant and resonant case. In non-resonant case, the tunable phase gate can be realized readily, while in resonant case a $\pi$-phase gate can be achieved after a full cycle of Rabi oscillation where the gate speed is much faster than that shown in previous schemes~\cite{20,24}. Moreover, the proposed scheme can be easily generalized to realize $N$-qubit phase gate. Discussion about the effect of possible experimental parameter errors on the fidelity of the entangled state are also given. Meanwhile, the influence of various decoherence processes such as the decay of the resonator modes, and the relaxation of the qudits is also investigated. Numerical results have shown a high fidelity to complete the phase gate.

\section*{Acknowledgments}

We greatly appreciate the support from the National Natural Science Foundation of China (NSFC)(No. 61604045).



\begin{thebibliography}{0}    
\bibitem{Shor} P.W. Shor, ``Polynomial-Time Algorithms for Prime Factorization and Discrete Logarithms on a Quantum Computer,'' SIAM J. Sci. Statist. Comput. {\bf 26}, 1484–1509 (1997).
\bibitem{Grover} L.K. Grover, ``Quantum Computers Can Search Rapidly by Using Almost Any Transformation,'' Phys. Rev. Lett. {\bf 80}, 4329 (1998).
\bibitem{3} J. Chiaverini, D. Leibfried, T. Schaetz, M.D. Barrett, R.B. Blakestad, J. Britton, W.M. Itano, J.D. Jost, E. Knill, C. Langer, R. Ozeri, D.J. Wineland, ``Realization of quantum error correction,'' Nature (London) {\bf 432}, 602 (2004).
\bibitem{4} M.S. Zubairy, A.B. Matsko, M.O. Scully, ``Resonant enhancement of high-order optical nonlinearities based on atomic coherence,'' Phys. Rev. A {\bf 65}, 043804 (2002).
\bibitem{5} L. M. K. Vandersypen, M. Steffen, G. Breyta, C. S. Yannoni, M. H. Sherwood, and I. L. Chuang, ``Experimental realization of Shor's quantum factoring algorithm using nuclear magnetic resonance,'' Nature (London) {\bf 414}, 883–887 (2001).
\bibitem{6} Y. S. Weinstein, M. A. Pravia, E. M. Fortunato, S. Lloyd, and D. G. Cory, ``Implementation of the Quantum Fourier Transform,'' Phys. Rev. Lett. {\bf 86}, 1889 (2001).
\bibitem{7} M. D. Reed, L. DiCarlo, S. E. Nigg, L. Sun, L. Frunzio, S. M. Girvin and R. J. Schoelkopf, ``Realization of three-qubit quantum error correction with superconducting circuits,'' Nature (London) {\bf 482}, 382-385 (2012).
\bibitem{8} L. Tornberg, M. Wallquist, G. Johansson, ,  V. S. Shumeiko, and G. Wendin, ``Implementation of the three-qubit phase-flip error correction code with superconducting qubits,'' Phys. Rev. B {\bf 77}, 214528 (2008).
\bibitem{9} D. G. Cory, M. D. Price, W. Maas, E. Knill, R. Laflamme, W. H. Zurek, T. F. Havel, and S. S. Somaroo, ``Experimental Quantum Error Correction,'' Phys. Rev. Lett. {\bf 81}, 2152 (1998).
\bibitem{10} C.-Y. Chen and S.-H. Li, ``Toffoli gate made from a single resonant interaction with a trapped ion system,'' Eur. Phys. J. D {\bf 41}, 557 (2007).
\bibitem{11} T. C. Ralph, K. J. Resch, and A. Gilchrist, ``Efficient Toffoli gates using qudits,'' Phys. Rev. A 75, 022313 (2007).
\bibitem{12} V. M. Stojanovi$\acute{c}$, A. Fedorov, A. Wallraff, and C. Bruder, ``Quantum-control approach to realizing a Toffoli gate in circuit QED,'' Phys. Rev. B {\bf 85}, 054504 (2012).
\bibitem{13} A. M. Chen, S. Y. Cho, and M. D. Kim, ``Implementation of a three-qubit Toffoli gate in a single step,'' Phys. Rev. A {\bf 85}, 032326 (2012).
\bibitem{14} X. Q. Shao, T. Y. Zheng, and S. Zhang, ``Robust Toffoli gate originating from Stark shifts,'' J. Opt. Soc. Am. B {\bf 29}, 1203–1207 (2012).
\bibitem{15} S. B. Zheng, ``Implementation of Toffoli gates with a single asymmetric Heisenberg XY interaction,'' Phys. Rev. A {\bf 87}, 042318 (2013).
\bibitem{16} B. P. Lanyon et al., ``Simplifying quantum logic using higher-dimensional Hilbert spaces,'' Nature Physics 5, 134-140 (2009). 
\bibitem{17} T. Monz, K. Kim, W. H$\ddot{a}$nsel, M. Riebe, A. S. Villar, P. Schindler,
M. Chwalla, M. Hennrich, and R. Blatt, ``Realization of the Quantum Toffoli Gate with Trapped Ions,'' Phys. Rev. Lett. {\bf 102}, 040501 (2009).
\bibitem{18} A. Fedorov, L. Steffen, M. Baur, M. P. da Silva, and A. Wallraff, ``Implementation of a Toffoli gate with superconducting circuits,'' Nature (London) {\bf 481}, 170 (2012).
\bibitem{19} A. Barenco, C. H. Bennett, R. Cleve, D. P. DiVincenzo, N. Margolus, P. Shor, T. Sleator, J. A. Smolin, and H. Weinfurter, ``Elementary gates for quantum computation,'' Phys. Rev. A {\bf 52}, 3457 (1995).
\bibitem{20} C. P. Yang, and S. Han, ``n-qubit-controlled phase gate with superconducting quantum-interference devices coupled to a resonator,'' Phys. Rev. A {\bf 72}, 032311 (2005).
\bibitem{21} L. M. Duan, B. Wang, and H. J. Kimble, ``Robust quantum gates on neutral atoms with cavity-assisted photon scattering,'' Phys. Rev. A {\bf 72}, 032333 (2005).
\bibitem{22} A. G$\acute{a}$bris and G. S. Agarwal, ``Vacuum-induced Stark shifts for quantum logic using a collective system in a high-quality dispersive cavity,'' Phys. Rev. A {\bf 71}, 052316 (2005).
\bibitem{23} X. Zou, Y. Dong, and G. C. Guo, ``Implementing a conditional 
z gate by a combination of resonant interaction and quantum interference,'' Phys. Rev. A {\bf 74}, 032325 (2006).
\bibitem{24} Y. Q. Zhang, S. Zhang, K. H. Yeon, and S. C. Yu, ``One-step implementation of a multiqubit controlled- phase gate with superconducting quantum interference devices coupled to a resonator,'' J. Opt. Soc. Am. B {\bf 29}, 300-304 (2012).
\bibitem{25} C. P. Yang, Y. X. Liu, and F. Nori, ``Phase gate of one qubit simultaneously controlling n qubits in a cavity,'' Phys. Rev. A {\bf 81}, 062323 (2010).
\bibitem{26} C. P. Yang, S. B. Zheng, and F. Nori, ``Multiqubit tunable phase gate of one qubit simultaneously controlling n qubits in a cavity,'' Phys. Rev. A {\bf 82}, 062326 (2010).
\bibitem{27} C. P. Yang, Q. P. Su, and J. M. Liu, ``Proposal for realizing a multiqubit tunable phase gate of one qubit simultaneously controlling n target qubits using cavity QED,'' Phys. Rev. A {\bf 86}, 024301 (2012).
\bibitem{28} C. P. Yang, Q. P. Su, F. Y. Zhang, and S. B. Zheng, ``Single-step implementation of a multiple-target-qubit controlled phase gate without need of classical pulses,'' Opt. Lett. {\bf 39}, 3312 (2014).
\bibitem{29} W. A. Li and G. Y. Huang, ``Deterministic generation of a three-dimensional entangled state via quantum Zeno dynamics,'' Phys. Rev. A {\bf 83}, 022322 (2011).
\bibitem{30} X. Q. Shao, L. Chen, S. Zhang, and K. H. Yeon, ``Fast CNOT gate via quantum Zeno dynamics,'' J. Phys. B: At. Mol. Opt. Phys. {\bf 42}, 165507 (2009).
\bibitem{31} X. Q. Shao, H. F. Wang, L. Chen, S. Zhang, Y. F. Zhao, and K. H. Yeon, ``One-step implementation of the 1 $\rightarrow$ 3 orbital state quantum cloning machine via quantum Zeno dynamics,'' Phys. Rev. A {\bf 80}, 062323 (2009).
\bibitem{32} A. Beige, D. Braun, B. Tregenna, and P. L. Knight, ``Quantum Computing Using Dissipation to Remain in a Decoherence-Free Subspace,'' Phys. Rev. Lett. {\bf 85}, 1762 (2000).
\bibitem{33} J. D. Franson, B. C. Jacobs, and T. B. Pittman, ``Quantum computing using single photons and the Zeno effect,'' Phys. Rev. A {\bf 70}, 062302 (2004).
\bibitem{flux1} V. E. Manucharyan, J. Koch, L. I. Glazman, and M. H. Devoret, ``Fluxonium: Single Cooper-Pair Circuit Free of Charge Offsets,'' Science {\bf 326}, 113-116 (2009).
\bibitem{flux2} G. Zhu, D. G. Ferguson, V. E. Manucharyan, and J. Koch, ``Circuit QED with fluxonium qubits: Theory of the dispersive regime,'' Phys. Rev. B {\bf 87}, 024510 (2013).
\bibitem{kim} M. D. Kim and J. Kim, ``Coupling qubits in circuit-QED cavities connected by a bridge qubit,'' Phys. Rev. A {\bf 93}, 012321 (2016).
\bibitem{34} D. I. Schuster, A. P. Sears, E. Ginossar, L. DiCarlo, L. Frunzio, J. J. L. Morton, H. Wu, G. A. D. Briggs, B. B. Buckley, D. D. Awschalom, and R. J. Schoelkopf, ``High-Cooperativity Coupling of Electron-Spin Ensembles to Superconducting Cavities,'' Phys. Rev. Lett. {\bf 105}, 140501 (2010).
\bibitem{35} R. Barends, J. Kelly, A. Megrant, D. Sank, E. Jeffrey, Y. Chen, Y. Yin, B. Chiaro, J. Mutus, C. Neill, P. O’Malley, P. Roushan, J. Wenner, T. C. White, A. N. Cleland, and John M. Martinis, ``Coherent Josephson Qubit Suitable for Scalable Quantum Integrated Circuits,'' Phys. Rev. Lett. {\bf 111}, 080502 (2013).

\end{thebibliography}
\end{document}